\documentclass[aps,showpacs]{revtex4-2}

\usepackage{color,url,hyperref}
\usepackage{graphicx}

\begin{document}

\title{
  Intrinsic coercivity induced by
  valence fluctuations in
  $4f$-$3d$ intermetallic magnets
}

\author{Hiroaki~Shishido$^1$}
\altaffiliation{shishido@pe.osakafu-u.ac.jp}
\author{Tetsuro Ueno$^2$}
\author{Kotaro Saito$^3$}
\author{Masahiro Sawada$^4$}
\author{Munehisa Matsumoto$^5$}
\altaffiliation[Present address: ]{Institute of Materials Structure Science, High Energy Accelerator Research Organization (KEK), 1-1 Oho, Tsukuba, Ibaraki 305-0801, Japan}

\affiliation{
$^1$Department of Physics and Electronics, Osaka Prefecture University, 1-1 Gakuen-cho, Naka-ku, Sakai, Osaka 599-8531, Japan\\
$^2$Quantum Beam Science Research Directorate, National Institutes for Quantum and Radiological Science and Technology, 1-1-1 Kouto, Sayo, Hyogo 679-5148, Japan\\
$^3$Institute of Materials Structure Science, High Energy Accelerator Research Organization, 1-1 Oho, Tsukuba, Ibaraki 305-0801, Japan\\
$^4$Hiroshima Synchrotron Radiation Center, Hiroshima University, Higashi-Hiroshima, Hiroshima 739-0046, Japan\\
$^5$Institute for Solid State Physics (ISSP), University of Tokyo, Kashiwa 277-8581, Japan}

\date{\today}

\begin{abstract}
Temperature dependence of magnetization curves of well homogenized samples of Ce(Co$_{1-x}$Cu$_{x}$)$_5$ ($0\le x \le 0.7$), a family of representative $4f$-$3d$ intermetallic magnets found in rare-earth permanent magnets, is measured.
A remarkable enhancement of intrinsic coercivity is observed with $x=0.3$ and $x=0.4$, persisting to higher temperatures.
This experimental observation is theoretically attributed to an effect of electronic correlation
 among $4f$-electrons. 
That is, an intrinsic pinning happens originating in an anomalously enhanced magnetic anisotropy energy contributed by an order of magnitude stronger charge-transfer process between $4f$-electrons and $3d$-electrons, than the conventional crystal field effects.
It is demonstrated that the $4f$-$3d$ charge-transfer process depends on the direction of magnetization in the middle of a crossover of the valence state of Ce between CeCu$_5$ with robust Ce$^{3+}$ and CeCo$_5$ with the mixed valence state.
\end{abstract}


%
%
%
%

\maketitle

\paragraph{Introduction}
Rare earth permanent magnets (REPMs) make a particular research target where the forefronts of fundamental physics meet the critical demands from society and industry~\cite{coey_2012}.
Among the intrinsic magnetism required for permanent magnets, magnetization and Curie temperature carried mostly by $3d$-electrons of Fe-group elements involve the long-standing problem of metallic magnetism~\cite{kubler}, and magnetic anisotropy is realized dominantly by $4f$-electrons, typically as an outcome from combined effects of crystal fields and spin-orbit interaction (SOI). 
SOI brings about novel topologically nontrivial electronic states with which novel functionalities are extensively investigated in the past decade~\cite{topo}, and has been one of the central topics in recent condensed matter physics.
Also, the multiple-scale problem to bridge over the gap between the microscopic magnetism and the macroscopic properties of bulk materials is not yet fully solved~\cite{stiffness_the_original,stiffness}.
Given the recent shifts to electric vehicles and introduction of robotics in various aspects of daily lives and industry, permanent magnets will be one of the most critically important materials in the upcoming decades. It is an important problem in many respects to construct the fundamental understanding of permanent magnets.

In this work, we point to a potential utility of Ce, which is relatively abundant among all of the rare earth elements~\cite{oliver_2018}, exploiting the valence fluctuations coming from the $4f$-electrons that reside on the border of localization and delocalization~\cite{smith_1982}.
While the typical magnetic anisotropy energy is understood to be driven by potential energy of well localized $4f$-electrons in the crystal fields, we argue that valence-fluctuations combined with SOI bring about an effective magnetic anisotropy driven by kinetic energy in the hybridization of delocalized $4f$-electrons and other conduction electrons. This kinetic-energy-driven anisotropy originating in charge fluctuations has an order of magnitude stronger energy scale than the magnetic anisotropy driven by potential energy.
Also the relatively strong coupling between $4f$-$3d$ electrons helps to render such kinetic-energy driven anisotropy more resistant against high temperature fluctuations~\cite{mm_2016}.

We present a case study with Ce(Co,Cu)$_5$.
The prototype, CaCu$_5$-type structure, (For a picture, see the supplementary material~\cite{supple}.)
makes one of the most fundamental building blocks among typical compounds in REPMs such as R$_2$T$_{17}$, RT$_{12}$, and R$_2$T$_{14}$B (R=rare earth and T=transition metals )~\cite{liandcoey_1991, rmp_1991, cromer_1959}.
Thus, the physics revealed with RT$_5$ should lead to a versatile guideline of material design for REPMs.
In Ce(Co,Cu)$_5$, there is a crossover in the valence state of Ce between CeCo$_5$ with mixed valence state of Ce~\cite{1982} and CeCu$_5$ with trivalent Ce~\cite{bauer_1987,bauer_1991}.
Between $3d$-electron ferromagnetism of CeCo$_5$ and $4f$-electron antiferromagnetism of CeCu$_5$~\cite{bauer_1987}, we observe the enhanced coercivity  and attribute it to the effects of valence fluctuations combined with SOI.
The whole composition range was investigated in Refs.~\cite{girodin_1985,meyer-liautaud_1987}  and magnetism trends in the Cu-rich compounds were investigated~\cite{mm_2020}, while positive contribution of valence fluctuations to the intrinsic magnetism has not been remarked so far~\cite{old_and_new}.

The materials RCo$_5$ (R=rare earth) with Co partly substituted by Cu belonged to the earliest REPMs ~\cite{strnat_1967,nesbitt_1968,tawara_1968} developed in 1960's. This eventually lead to the Sm-Co-based magnets that were the champion magnet in 1970's before the advent of Nd-Fe-B magnets~\cite{sagawa_1984,croat_1984}.
While the main-stream magnets attracted attention both technologically and fundamentally since 1980's~\cite{rmp_1991}, the understanding of the early REPMs was not completely established.
The fundamental understanding of the Nd-Fe-B magnets are still being constructed~\cite{hono_2012,stiffness_the_original}.
In these respects, all possible roles of $4f$-electron physics in the $3d$-electron ferromagnetism are worth investigating comprehensively. 
Valence fluctuations characterize some of the important rare earth elements such as Ce and Sm.

\paragraph{Experimental setup}
Poly-crystalline samples of Ce(Co$_{1-x}$Cu$_x$)$_5$ for $x=0$, $0.3$, $0.4$, $0.5$, $0.6$, and $0.7$ were grown by arc-melting at Ar atmosphere from elementary materials of 3N(99.9\%)-Ce, 3N-Co and 4N-Cu.
They were wrapped in Ta-foil, sealed in a quart tube under vacuum and annealed at 850-1050$^\circ$C for 5-24 days according to each concentration.
The samples were characterized by X-ray diffraction~(XRD), scanning electron microscope~(SEM) and energy dispersive X-ray spectrometry (EDX). 
XRD measurements were performed by a SmartLab (Rigaku) with Cu-K$\alpha$ radiation.
SEM images and EDX mapping were collected with a SU8010~(Hitachi) and XMax50 EDX (Oxford instruments).
Magnetization measurements were carried out with a MPMS3 (Quantum design) for the magnetic field range of $-70$\,kOe to $70$\,kOe in the temperature range of $1.9$\,K to $300$\,K.
X-ray absorption spectroscopy (XAS) experiment was performed at the HiSOR BL-14 at the Hiroshima Synchrotron Radiation Center (HSRC), Hiroshima University, Japan~\cite{Sawada2007}.
The details of the experimental station are described in Refs.~\onlinecite{Sawada2010} and~\onlinecite{Ueno2012}.
Sample was filed by a diamond file in the ultra high vacuum (UHV) to obtain clean surface.
Ce $M_{4,5}$ XAS spectra were measured using the total electron yield method by measuring the sample drain current.
The sample temperature was maintained during the measurements using a closed-cycle He refrigerator.

\begin{figure}
\includegraphics[width=8cm]{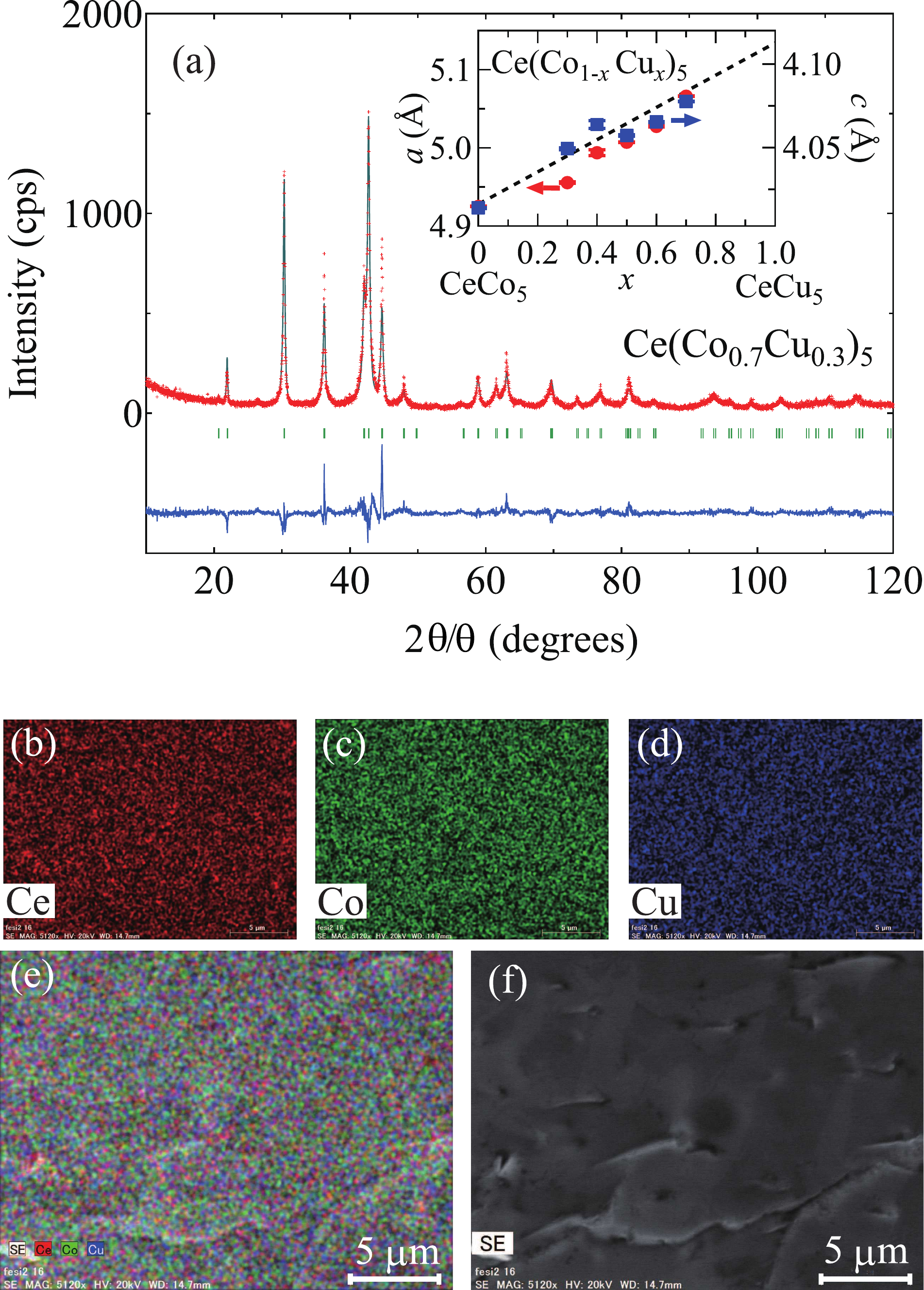}
\caption{(Color online)
  (a)  The powder diffraction pattern for Ce(Co$_{0.7}$Cu$_{0.3}$)$_5$.
  Red dots, black and blue lines denote the experimental data, fitting results and their difference. SEM-EDX chemical mapping of Ce(Co$_{0.7}$Cu$_{0.3}$)$_5$ for (b) Ce, (c) Co, (d) Cu, and (e) overlay of them on the SEM image. (f) SEM image corresponding to the SEM-EDX mapping.}
\label{fig::XRD_SEMEDX}
\end{figure}

\paragraph{Characterization of the samples}
Main panel of Fig.~\ref{fig::XRD_SEMEDX}(a) shows the powder XRD pattern for Ce(Co$_{0.7}$Cu$_{0.3}$)$_5$.
All peaks are well indexed to belong to the CaCu$_5$-type phase (Space group: $P6/mmm$ \#191, and no contamination was detected within the experimental errors. 
Rietveld XRD refinements were carried out by RIETAN-FP~\cite{Izumi} for the samples. 
Obtained lattice constants $a$ and $c$ follow the Vegard's law as shown in the inset of Fig.~\ref{fig::XRD_SEMEDX}(a).
It indicates no significant phase separation occurs even in intermediate Cu concentrations.
Figures~\ref{fig::XRD_SEMEDX}(b)--(f) show a typical SEM image (Fig.~\ref{fig::XRD_SEMEDX}(f)) and SEM-EDX mapping (Figs.~\ref{fig::XRD_SEMEDX}(b)--(e)) of Ce(Co$_{0.7}$Cu$_{0.3}$)$_5$.
SEM images of Ce(Co$_{1-x}$Cu$_{x}$)$_5$ shows no particular microstructure that may work as an extrinsic pinning center.
Moreover, chemical distributions of Ce(Co$_{1-x}$Cu$_{x}$)$_5$ are homogeneous as shown in Figs.~\ref{fig::XRD_SEMEDX} (b)--(e).
Considering XRD and SEM-EDX results, we can conclude that the samples are single phase and homogeneous. 
It implies that the observed bulk properties presented below can be attributed to the intrinsic properties.

\begin{figure}
  \includegraphics[width=8cm]{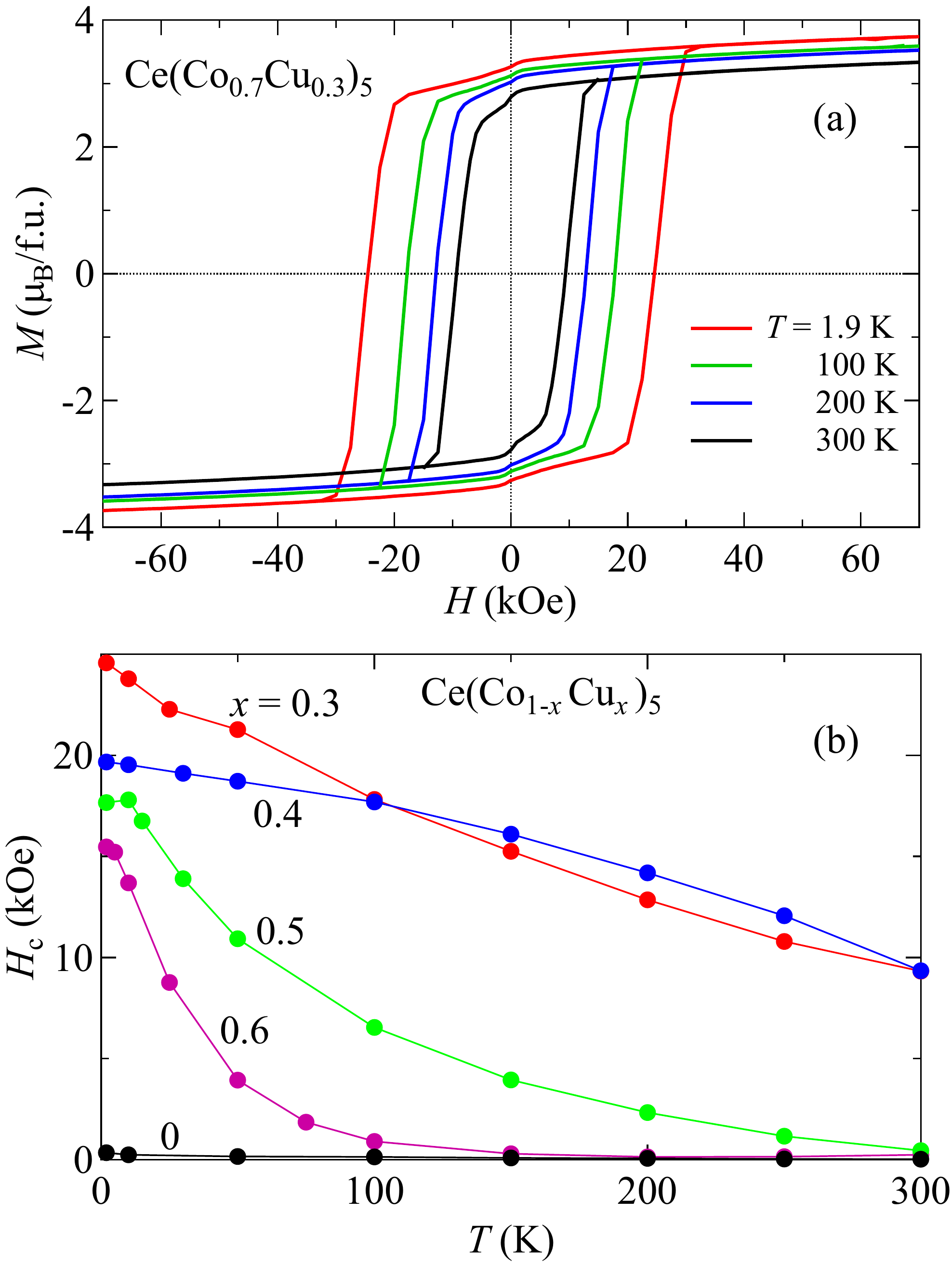}
  \caption{(Color online)
      (a) Magnetization curves of Ce(Co$_{0.7}$Cu$_{0.3}$)$_5$ at various temperatures and (b) Temperature dependence of coercive force $H_{\rm c}$ for various substitutions of Ce(Co$_{1-x}$Cu$_{x}$)$_5$.}
\label{fig::MH} 
\end{figure}
\paragraph{Results: measurement of bulk magnetization curves}
Reflecting a ferromagnetic ordering,
magnetization curves of Ce(Co$_{0.7}$Cu$_{0.3}$)$_5$ exhibit hysteresis loops, as shown in Fig.~\ref{fig::MH}(a).
Hysteresis loops show abrupt drop across zero magnetic field.
This anomaly becomes more pronounced at $x=0.4$, and hysteresis loops show snake-like form at $x=0.5$ and $0.6$ (shown in the Supplementary Material~\cite{supple}).
The origin of this anomalous behavior is not clear yet, but finite concentration distributions could provide negligible difference on coercive force $H_{\rm c}$ and saturated magnetization ($M_{\rm s}$), and the combination of different hysteresis loops gives these anomalous loops at intermediate concentration region. 
For Ce(Co$_{0.7}$Cu$_{0.3}$)$_5$, $H_{\rm c}$ of 9\,kOe increases with decreasing temperature, and reaches 24.6\,kOe at 1.9\,K.
Temperature dependence of $H_{\rm c}$ for each substitution are summarized in Fig.~\ref{fig::MH}~(b).
The measured
$H_{\rm c}$ for CeCo$_5$ is almost temperature independent and weak, indicating that CeCo$_5$ is effectively a soft magnetic material~\cite{footnote1}.

The measured $H_{\rm c}$ at 1.9\,K has a maximum at $x= 0.3$, decreases almost linearly with increasing $x$.
More interestingly, $H_{\rm c}(T)$ for $x = 0.5$ and $0.6$ exhibit a downward convex temperature dependence, whereas that for $x=0.4$ shows a slightly upward convex temperature dependence.
The temperature dependence of $H_{\rm c}(T)$ with $x=0.3$ is almost linear.
Such significant enhancement of the temperature resistance in $H_{\rm c}(T)$ around $x=0.3$ and $x=0.4$ should come from the direct hybridization between $4f$-electron state and the conduction electron states that include the ferromagnetic $3d$-electron state of Co, in contrast to the weak temperature resistance in the materials with the indirect exchange coupling between well localized $4f$-electrons in Nd and $3d$-electrons~\cite{mm_2016}.

\paragraph{Microscopic probing of the Ce valence state}
\begin{figure}
  \includegraphics[width=8cm]{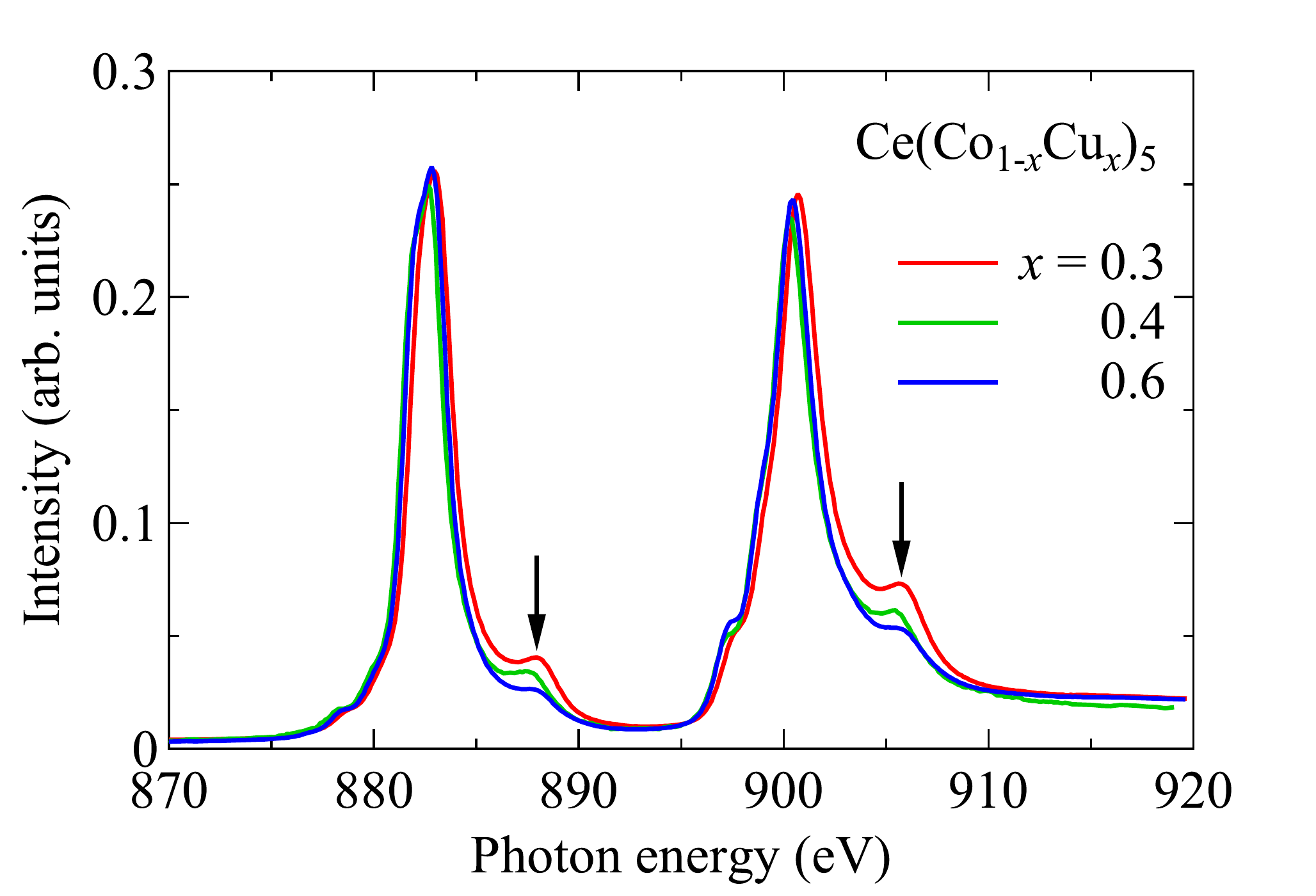}
  \caption{(Color online)
      Soft X-ray absorption spectra of Ce(Co$_{1-x}$Cu$_{x}$)$_5$ ($x=0.3$, $0.4$, and $0.6$) at 20~K.
      The arrows indicate the presence of the contribution from the $(4f)^0$ state of Ce$^{4+}$ in the target materials as the initial state of the core-electron excitation~\cite{fuggle_1983}, which is gradually suppressed from $x=0.3$ to $x=0.6$.}
\label{fig::XAS}
\end{figure}
The measured XAS results at $20~\mbox{K}$ for various $x$ are shown in Fig.~\ref{fig::XAS}.
The absorption peak corresponding to the $(4f)^0$ state of Ce$^{4+}$ in the target materials is located at around 5~eV higher from the main peak originating in the $(4f)^1$ state of Ce$^{3+}$.
Here we refer to Ref.~\onlinecite{fuggle_1983} for the assignment of the peaks.
We focus on the relative trends among the samples concerning the contribution from the Ce$^{4+}$ state and we see that the measured data clearly shows the trend of the valence fluctuations being suppressed as the concentration of Cu increases.
Here we note that $T$-dependence of XAS was relatively minor as compared to the $x$-dependence~\cite{supple}.

\paragraph{A supporting data set of electronic structure calculations}
The crossover in the valence state of Ce by Cu substitution in CeCo$_5$ can be simulated by imposing Coulomb $U_{4f}$ among $4f$-electrons.
The realistic range of the parameter $U_{4f}$ is upper-bounded by the localized limit at $U_{4f}=5$~eV.
Between such localized limit and the mixed valence region with $U_{4f}\ll 5$~eV to which the experimental pristine CeCo$_5$ seems to correspond, the crossover in the valence state of Ce can be numerically observed as the parameter $U_{4f}$ is varied starting from $U_{4f}=0$. 
In this way, we below present a set of electronic structure calculations based on the density functional theory (DFT)~\cite{HK_1964} combined with the $U_{4f}$ parameter for CeCo$_5$, so called the DFT+U method~\cite{anisimov_1991}, with which we describe the trend in the valence state of Ce in Ce(Co$_{1-x}$Cu$_x$)$_5$ from $x=0$ to $x$ closer to $1$.

We use the DFT code, {\it OpenMX}~\cite{Ozaki2003,Ozaki2004,Ozaki2005,Duy2014,Lejaeghere2016}, based on a linear combination of pseudoatomic orbitals.
The DFT+U method and the SOI works on the $j$-dependent pseudopotential~\cite{MBK1993,Theurich2001} for the description of the non-collinear magnetism within DFT (Here $j$ refers to the total angular momentum incorporating both of spin and orbital). 
We proceed with the following two steps:
a) starting with the experimental lattice constants, the crystal structure is computationally optimized, and then
b) the intrinsic magnetic anisotropy is estimated by the calculation of the energy while putting a constraint on the direction of magnetization. 
Here the SOI is turned on for b). 
Calculated energy is plotted with respect to the angle $\theta$ of the direction of magnetization as measured from the $c$-axis.
The particular DFT+U scheme we use here is based on the formulation in Ref.~\onlinecite{mj_2006} as implemented in {\it OpenMX}.
We set another parameter $U_{3d}=2.0$~eV for Co in order to describe a set of realistic scales in the $3d$-electron part~\cite{supple}.

\begin{figure}
\includegraphics[width=12cm]{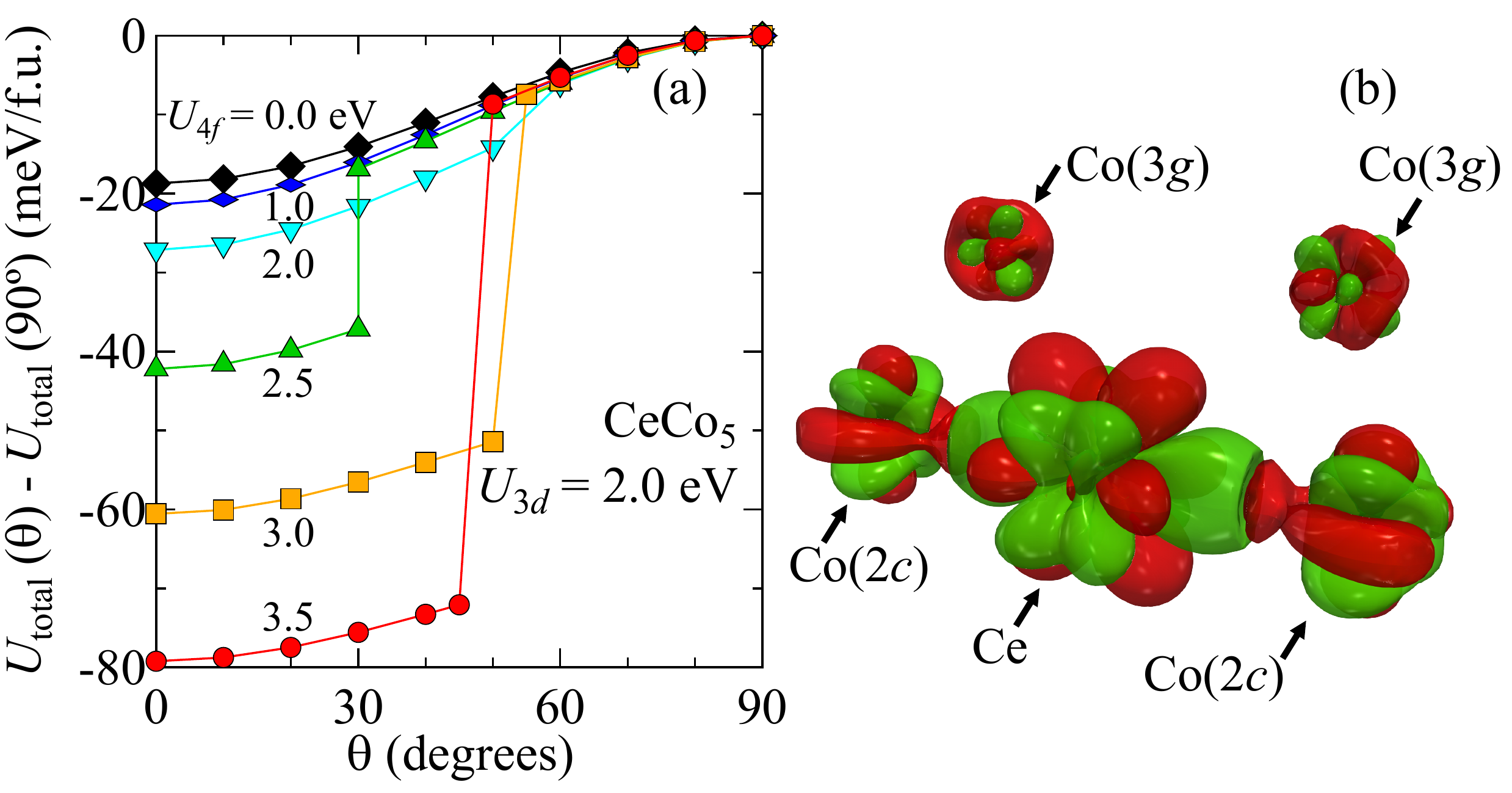}
\caption{\label{fig::results_E_vs_theta}
(Color online)
(a) Calculated energy as a function of $\theta$, the angle between the magnetization and the c-axis, for CeCo$_{5}$ within GGA+U. The parameters in the calculations are $0\le U_{4f}\le 3.5~\mbox{eV}$ and $U_{3d}=2.0~\mbox{eV}$.
(b) Visualized difference of the calculated spin density between $\theta=10^\circ$ and $\theta=90^\circ$ with $U_{4f} = 3.5~\mbox{eV}$ and $U_{3d}=2.0~\mbox{eV}$.
      The red (green) region corresponds to the area where the spin density has increased (decreased), respectively. The path of a charge transfer process from the Ce atom to the Co($2c$) atom can be seen.
}
\end{figure}
Calculated energy $U_{\rm total}$ as a function of $\theta$ is shown in Fig.~\ref{fig::results_E_vs_theta}(a). 
We observe the enhancement of the effective magnetic anisotropy energy from $U_{4f}=2$~eV to $U_{4f}=3.5$~eV, gradually showing a larger jump in the calculated energy.
We observed that this type of jump at $\theta_{\rm jump}$ is encountered irrespective of the ways the particular angle $\theta_{\rm jump}$ is approached.
The overall calculations proceed as close as possible to the conventional calculations of magnetic anisotriopy energy, always starting from a converged electronic state in the crystal structure optimization in 
a) above and then imposing the constraint on the direction of magnetization in b). 
When the convergence is not reached in b) at some $\theta$, other converged electronic state with a neighboring value of $\theta$ is taken as a starting point.
In this way, the jump in the calculated energy as a function of $\theta$ is robustly observed. Accordingly enhanced effective magnetic anisotropy energy does not seem to be an artifact of our numerics. 

Furthermore, we can attribute the observed jump in $U_{\rm total}(\theta)$ to the valence fluctuation and the associated charge transfer between Ce and Co, using a visualization of the electronic state in a convincing way as follows.
For the calculated electronic structure with $U_{4f}=3.5$~eV, difference of the calculated spin density is taken between $\theta=10${$^\circ$} and $\theta=90${$^\circ$} and its real-space picture is shown in Fig.~\ref{fig::results_E_vs_theta}(b).
A charge transfer path between Ce and Co($2c$) is observed.
This indicates that the jump in the calculated energy originates in the delocalization of $4f$-electron in Ce, which hops to the neighboring Co($2c$) atoms, and the magnetic anisotropy has been enhanced by the charge transfer energy. 
This particular direction-dependent $4f$-$3d$ charge transfer process combined with spin-orbit interaction contributes to the anomalous enhancement of the intrinsic magnetic anisotropy. 
Since the energy scale of $4f$-$3d$ hybridization in Ce is $O(0.1)$~eV~\cite{fuggle_1983} and the scales of the magnetic anisotropy energy originating in the crystal fields is of the order of $O(10)$~meV, the effective magnetic anisotropy can be enhanced by an order of magnitude by the involvement of the valence fluctuations. 
With such enhancement of the effective uni-axial anisotropy $K$ and the reduction of magnetization as realized by the dilution effect caused by Cu into ferromagnetic Co, the domain wall width spanning the length scale of $\pi\sqrt{A/K}$ (here $A$ is the exchange stiffness involving magnetization)~\cite{kittel_1949}, which is in the order of $O(1)$~nm for pristine CeCo$_5$, can narrow down to the scale of $O(1)$~\AA.
Then an atomic-scale pinning of the domain walls can be at work.
With the measurements of our homogenized samples and the supporting calculations, this mechanism of an intrinsic coercivity seems to be plausible.

\paragraph{Discussions: implications on Sm compounds}
The relevance of valence fluctuations for the anomalous magnetic anisotropy may hold for other rare earth elements with valence fluctuations, such as Sm, Eu, Tm, Yb, and Pr under pressure, and for early actinides such as U, Np, and Pu, where the $5f$-electrons located on the border of localization and delocalization~\cite{smith_1982}.
A possible close example can happen with Sm-based compounds.
For another representative $4f$-$3d$ intermetallic ferromagnet for REPMs, Sm$_2$Fe$_{17}$N$_3$, an electronic structure calculation pointed to a divalent state of Sm in the ground state~\cite{ogura_2015}.
In contrast, the XAS data of the same compound and related ones at finite temperatures indicate that Sm is in the trivalent state~\cite{Ueno_SmFeCo}. Even though some sample dependence can be suspected~\cite{review_1995}, these past claims point to the possibility of a valence transition or a crossover for Sm at very low temperatures.
Common physics with Ce(Co,Cu)$_5$ can enhance the magnetic anisotropy of Sm-based Fe-group intermetallics, which may be at work in Sm(Co,Cu)$_5$ in the early days of REPM~\cite{nesbitt_1968}, where an enhancement of the coercivity was reported for the intermediate region in the concentration of Cu.

\paragraph{Discussions: materials design for Fe-rich cases}
Since CeFe$_5$ does not exist as a bulk stable material and it often happens that Cu does not mix with Fe, we have not explicitly addressed the Fe-rich cases.
However, the physics with the delocalized $4f$-electrons of Ce in the ferromagnetically polarized conduction bands should generically hold:  the Fe-rich Ce intermetallics and their less ferromagnetic counterparts that may be realized by mixing Fe with Al or Ga, instead of Cu, should show an analogously enhanced anisotropy induced by valence fluctuations. This may be exploited in order to implement an intrinsic coercivity in the possible fabrication of a rare-earth permanent magnet based on Fe-rich $4f$-$3d$ intermetallics.

\paragraph{Conclusions and outlook}
We have described how the valence fluctuations of Ce can be relevant for the enhanced magnetic anisotropy and its associated intrinsic coercivity with the case of Ce(Co$_{1-x}$Cu$_{x}$)$_5$, where the intrinsic coercivity is maximized around $x=0.3$.
This particular coercivity is characterized by its relatively weak temperature dependence due to the direct hybridization mechanism of the delocalized $4f$-electrons and conduction electrons.
The temperature resistance can pave the way toward the possible technological applications at elevated temperatures.

Apparently, the intrinsic coercivity associated with valence fluctuations has not been discussed so far.
This points to a possibility that there are potentially several mechanisms of the coercivity of permanent magnets.
There can be some more coercivity mechanisms that are not well explored but are relevant both for fundamental understanding and industrial applications. 
In order to construct a versatile understanding of complicated materials such as REPMs, spanning from the Nd-Fe-B magnets~\cite{sagawa_1984,croat_1984,rmp_1991} to Sm-Co magnets and possible new ones~\cite{byGeorge}, it is desirable to implement a more comprehensive way of material characterization utilizing multiple probes  and to fairly investigate a range of the possible mechanisms based on the fundamentals of statistical physics and condensed-matter physics.

\begin{acknowledgments}
Inspiring discussions with Y.~Kuramoto and introduction to the present problem given by T.~Akiya are gratefully acknowledged.
The present project was partly supported by JSPS KAKENHI Grant No.~15K13525 and a Hitachi Metals-Materials Science Foundation Research Fellowship.
The XAS experiments were performed at HSRC with the approval of the Proposal Assessing Committee (proposal Nos. 16AG045 and 17BG017).
Numerical calculations for the present work were done in the facilities in ISSP Supercomputer Center, University of Tokyo.
\end{acknowledgments}


\begin{thebibliography}{99}
\bibitem{footnote1}
Even with the moderately strong uni-axial magnetic anisotropy of the pristine CeCo$_5$,
that is not strong enough to let the material qualify as a hard magnetic material.
\end{thebibliography}
\end{document}